\documentclass[11pt, a4paper]{article}
\usepackage{jheppub}
\usepackage{graphicx}
\usepackage{epsfig,amsmath,amsthm}
\usepackage{subfig}
\newcommand{\be}{\begin{equation}}
\newcommand{\ee}{\end{equation}}
\newcommand{\bea}{\begin{eqnarray}}
\newcommand{\eea}{\end{eqnarray}}
\def\bse{\begin{subequations}}
\def\ese{\end{subequations}}

\def\IZ{\relax\ifmmode\hbox{Z\kern-.4em Z}\else{Z\kern-.4em Z}\fi}

\newcommand{\non}{\nonumber \\}
\newcommand{\reef}[1]{(\ref{#1})}
\newcommand{\ie}{{\it i.e.,}\ }
\newcommand{\labell}[1]{\label{#1}}

\def\del{{\partial}}

\def\al{\alpha} \def\bt{\beta}

\def\bi{\begin{itemize}} \def\ei{\end{itemize}}

\def\({\left(} \def\){\right)}
\def\[{\left[} \def\]{\right]}

\begin{document}

\title{ \center{Modification of late time phase structure by quantum quenches}}

\author{Ling-Yan Hung$^1$, Michael Smolkin$^1$ and Evgeny Sorkin$^2$\\
{\it ${}^1$ Perimeter Institute for Theoretical Physics,
\\ 31 Caroline Street North, Waterloo,
Ontario N2L 2Y5, Canada} \\
{\it ${}^2$ Department of Physics and Astronomy, \\
University of British Columbia,\\
Vancouver, BC V6T 1Z1, Canada} \\

  {\tt\href{mailto:jhung@perimeterinstitute.ca}{jhung@perimeterinstitute.ca}}\\
 {\tt\href{mailto:rmyers@perimeterinstitute.ca}{msmolkin@perimeterinstitute.ca}}\\
 {\tt \href{mailto:evgeny@phas.ubc.ca}{evgeny@phas.ubc.ca}}

}

\begin{abstract}{The consequences of the sudden change in the coupling constants (quenches) on the phase structure of the theory at late times are explored. We study in detail the three dimensional $\phi^6$ model in the large $N$ limit, and show that
the $\phi^6$ coupling enjoys a widened range of stability compared to the static scenario.  Moreover, a new massive phase emerges, which for sufficiently large coupling becomes the dominant vacuum. We argue that these novel phenomena cannot be described by a simple thermalization effect or the
emergence of a single effective temperature. }
\end{abstract}

\maketitle

\section{Introduction}

Unlocking the problem of out-of-equilibrium dynamics of a quantum coherent system is one of the
fundamental questions in quantum phyiscs. This is particularly true in the context of quantum field theories,
where many important questions have so far been addressed
mainly
in a static scenario, such as the renormalization group flow, the
phase structure of vacuua and critical points.
There are also interesting questions that
spring directly from a non-equilibrium system, such as the mechanisms of relaxation, the time-scale over which this occurs,
and the existence of an effective description at late times.
These problems, given their fundamental role in field theories, naturally appear in many places.
For example, it is not surprising that non-equilibrium dynamics are important in cosmology, in the evolution of the early
universe.
The RHIC experiments, involving the relaxation of the quark-gluon plasma,
is another such example.
Dynamical systems
appear frequently in the context of condensed matter physics.
Recently, the study has been
rendered particularly pertinent experimentally due to new advances in the control of cold atomic gases\cite{relax,relax1,relax2,Newcrad}. For the first
time, we are able to observe minute details of the evolution of a system that retains its quantum coherence for
sufficiently
long periods of time.
One class of situations that has been subjected to intensive studies is called the \emph{quantum quench}, in
which a particular external field, or parameter, of the system is changed abruptly. An
example is a sudden change of the external magnetic field to which the atoms couple. These
experiments have inspired a flurry of theoretical activities, most
notably initiated by Calabrese and Cardy \cite{cardyquench}.
Previous works however, have been concentrated on free field theories,
one-dimensional interacting theories and integrable models,  see e.g. \cite{oned1,oned2,oned3,oned4,rigol,CEF, caneva}
Attempts to understand interacting theories in higher dimensions by considering
the large $N$ limit of a $\phi^4$ theory have been made in \cite{Sotiriadis:2010si}.(See also
the related problem \cite{Das:2012mt}).

Previous studies of
the quenches
concerned mostly about the relaxation of the system. One
central issue is whether the system thermalizes and
therefore describable
by an effective temperature
at late times. It is however an open
problem if thermalization occurs at all, and 
if it does not, which is shown to be the case in many integrable models
and in some cases even interacting models (see for example \cite{rigol}), whether there are  convenient
effective descriptions of such systems and observables or effective parameters that
characterize their behavior. 
This leads us to the current investigation of 
the \emph{phase structure} of some out-of-equilibrium state, which in the scenario concerned is prepared
by a quantum quench. This should be contrasted with the usual notion of the phase structure of a given
Hamiltonian, which is a property of its ground state. Here, we have to deal with a state, which, while
settling to some \emph{static equilibrium} in the far future, does not resemble a thermal state,
nor is it able to relax to the ground state because of its isolation and 
energy conservation after the quench. It is therefore only natural
to consider fluctuations about such a special state as oppose to the ground state,
and determine its corresponding \emph{phase structure}.

In this work, we demonstrate that this phase structure differs significantly from that
of the ground state even at late times as the system
approaches equilibrium again. 

In particular, we explore the $g_6\phi^6$ theory at the tricritical point, \ie when all dimensionful couplings except the physical mass immediately after the quench are tuned to zero. What is special about this model is that it was shown to possess an ultraviolet fixed point $g_6=192$ using the $1/N$ expansion about $N=\infty$ \cite{Townsend:1975kh, Pisarski:1982vz}. This fixed point however lies in the instability region of  the model where the non-perturbative effects dominate\footnote{See also \cite{Aharony:2011jz} for recent analysis of the $\beta$-function in the case of three dimensional Chern-Simons theories coupled to a scalar field in the fundamental representation.} \cite{Bardeen:1983rv}. The latter implies that the theory is always driven into the unstable region by the $\beta$-function and therefore apparently does not make physical sense.
We revisit this theory in the context of quantum quenches.
To that end, we employ the methods introduced in \cite{Sotiriadis:2010si},
where the effect of a quench is incorporated as a boundary condition on the fields.
Assuming that the system does settle down, we then self-consistently compute the
effective potential which defines the notion of phase structure 
of the theory at late times. 
We thus obtain a corresponding phase diagram, which surprisingly is 
modified dramatically in comparison to the unquenched case. 
The region of stability is substantially widened such that the UV fixed point of the $\beta$-function now lies well within.
Moreover, a new stable minimum in the effective potential emerges when the coupling constant exceeds the upper bound of the stability range in
the static theory. The new vacuum starts life as a meta-stable phase, but then becomes dominant for sufficiently large values of the coupling. In particular, the effective mass of the new phase increases as the coupling increases, and eventually diverges when the coupling hits the boundary of a newly established range of stability.

In the following,
we will employ large $N$ expansion and study in detail the quench dynamics of the massive scalar $O(N)$ vector model.

\section{Quenching the scalar model}
The scalar $O(N)$  vector model consists an $N$-component scalar field $\phi$.
For simplicity we assume that initially the theory is free and the system is prepared in the ground state of a free hamiltonian $| \Psi_0 \rangle$. At $t=0$ the marginal $\phi^6$ interaction as well as relevant $\phi^4$ interaction are instantaneously switched on, and at the same instant the bare mass parameter of the field jumps from $\mu_0$ to $\mu$. The action of the system after the quench is given by
 \be
 S(\phi)={1\over 2} \int d^3x \big[\del_\mu\phi\del^\mu\phi-\mu^2\phi^2-{g_4\over  2 N}(\phi^2)^2-{g_6\over  3 N^2}(\phi^2)^3\big]~.
 \labell{scalaraction}
 \ee

Since parameters of the theory are changed abruptly rather than adiabatically, one needs to resort to the well-known Keldysh-Schwinger, or in-in, formalism for non-equilibrium quantum systems. In this formalism the integration over time coordinate $t$ in the path integral starts from some initial time $t_i$, extends to some final time $t_f$ and then goes back to $t_i$. Correlation functions are path ordered. 
In this approach one needs to impose boundary conditions at $t=t_i$. In our case
we impose that the initial state at $t_i=0$ is given by $|\Psi_0\rangle$.

The expectation value of an arbitrary operator $\mathcal{\hat O}(t)$ is given by
 \be
 \langle \Psi_0 | \mathcal{ \hat O}(t)   |\Psi_0\rangle
 =\int_{CTP} D\phi \, \mathcal{ \hat O}(t) \, e^{i S(\phi)}~,
 \ee
where for brevity we use the following notation to designate the closed-time-path (CTP) integral measure
 \be
 \int_{CTP} D\phi=\int D\phi_i \, \Psi_0(\phi_i)\int D \tilde\phi_i  \, \Psi_0^*(\tilde\phi_i)
 \int_{\phi_i}^{\tilde\phi_i}D\phi ~, 
 \ee
where $\phi_i$ and $\tilde\phi_i$ denote the values of the scalar field $\phi$ at the end points of the time contour, whereas $\Psi_0(\phi_i)=\langle \phi_i|\Psi_0\rangle$ and similarly for the complex conjugate $\Psi_0^*(\tilde\phi_i)$.

Introducing the following identity into the path integral\footnote{We keep CTP label in the path integral over $\rho$ and $\lambda$ to emphasize that the delta-function is inserted at each point of the Keldysh-Schwinger contour. Obviously there are no boundary conditions associated with $\rho$ and $\lambda$.}
\be
\mathbb{I} \sim \int_{CTP} D\rho\,\delta(\phi^2-N\rho)\sim\int_{CTP} D\rho D\lambda ~e^{-{i\over 2}\int d^3x \,\lambda(\phi^2-N\rho)}~,
\label{identity}
\ee

yields
 \be
 \langle \Psi_0 | \mathcal{ \hat O}(t)   |\Psi_0\rangle
 = \int_{CTP} D\phi \int D\rho D\lambda\, \mathcal{ \hat O}(t) \, e^{i S(\phi,\rho,\lambda)}~,
 \ee
where
 \be
 S(\phi,\rho,\lambda)={1\over 2} \int d^3x \big[\del_\mu\phi\del^\mu\phi-(\mu^2+\lambda)\phi^2-N{g_4\rho^2\over  2}-N{g_6\rho^3\over  3}+N\rho\lambda\big]~.
 \ee
Performing now the Gaussian integral over $\phi$ leads to
 \be
 \langle \Psi_0 | \mathcal{ \hat O}(t)   |\Psi_0\rangle
 =\int_{CTP} D\rho D\lambda \, \mathcal{ \hat O}(t) \, e^{i N S_{eff}(\rho,\lambda)}~,
 \label{pathint}
 \ee
with
 \be
 S_{eff}(\rho,\lambda)=\int d^3x\[{\lambda\,\rho\over 2}-{g_4 \over 4}\rho^2-{g_6 \over 6}\rho^3\]
 +{i\over 2}\text{Tr}\log(\square+\mu^2+\lambda)~.
 \label{scalaction}
 \ee

The first thing to note about the above expression is that boundary conditions are now encoded in the functional trace. Secondly, this trace explicitly depends on the integration parameter $\lambda$, and this in turn renders evaluation of
the remaining path integral very difficult.

However, in the limit when $N$ is large while $g_4$ and $g_6$ are fixed,
the right hand side of \reef{pathint} is dominated by the field configurations which minimize \reef{scalaction}, \ie solutions of the corresponding classical equations of motion. The effective mass can thus be evaluated. This is often called
the stationary phase approximation. This gives
 \bea
 m_\phi^2&=&\mu^2+g_4\bar\rho+g_6\bar\rho^2~,
 \non
  \bar\rho&=&\int {d^2 p \over (2\pi)^2} \, \tilde G_\phi(t,t;p)~,
  \label{scalargap}
 \eea
where $m_\phi^2=\mu^2+\bar\lambda$ is the effective mass of the scalar field and $\tilde G_\phi(t_1,t_2;p)$ is
the full momentum space  two point correlation function of the scalar field to leading order in $1/N$. Fields evaluated at the saddle point are
denoted by a bar.

Note that $\tilde G_\phi(t_1,t_2;p)$ depends on the effective mass $m_\phi^2$, and therefore it is difficult to solve \reef{scalargap} in full generality. Hence, in what follows we use the approximation proposed in \cite{Sotiriadis:2010si}. In particular, we assume that $m_\phi$ tends to a stationary value $m_{\phi}^{*}$ and that this happens fast enough to be approximated by a jump.
Then the  two point correlation function $\tilde G_\phi(t_1,t_2;p)$ is approximately the same as the propagator in the
massive free field theory in which the physical mass is instantaneously changed from $\mu_0$ to $m_{\phi}^*$. \ie
 \be
 \tilde G_\phi(t_1,t_2;p)\simeq G_\phi(t_1,t_2;p;\mu_0, m^*_\phi)~,
 \ee
where \cite{Sotiriadis:2010si}
 \be
 G_\phi(t_1,t_2;p;\mu_0, m^*_\phi)={(\omega_p^*-\omega_{0p})^2\over 4\omega_p^{*2}\omega_{0p}}\cos\omega_p^*(t_1-t_2)
 +{\omega_p^{*2}-\omega_{0p}^2\over 4\omega_p^{*2}\omega_{0p}}\cos\omega_p^*(t_1+t_2)
 +{1\over 2\omega_p^*} e^{-i\omega_p^*|t_1-t_2|}~,
 \label{phicorr}
 \ee
with  $\omega_{p}^*=\sqrt{\vec p^{\,2}+m_\phi^{*2}}$ and  $\omega_{0p}=\sqrt{\vec p^{\,2}+\mu_0^2}$. The second term on the right hand side is the only one that breaks time translation invariance.
However, its contribution to $\bar\rho$ vanishes for $t_1=t_2=t\rightarrow\infty$ within our stationary phase approximation.
Therefore \reef{scalargap} yields the following equation for $m^*_\phi$
 \begin{multline}
 m_\phi^{*2}=\mu^2-{g_4\over 4\pi}\bigg(\mu_0+{1\over 2}\sqrt{m_\phi^{*2}-\mu_0^2}\,\arccos
 \bigg({\mu_0\over m_\phi^*}\bigg)-\Lambda\bigg)
 \\
 +{g_6\over 16\pi^2}\bigg(\mu_0+{1\over 2}\sqrt{m_\phi^{*2}-\mu_0^2}\,\arccos
 \bigg({\mu_0\over m_\phi^*}\bigg)-\Lambda\bigg)^2~,
 \end{multline}
where we have taken a sharp cut off $\Lambda$ to regulate the divergent integral over the momentum.
To eliminate the cut off dependence we apply the following renormalization scheme
 \bea
 \mu_R^2&=&\mu^2+{g_4\over 4\pi}\Lambda+{g_6\over 16\pi^2}\Lambda^2~,
 \non
 g_4^R&=&g_4+{g_6\over 2\pi}\Lambda~.
 \label{renparam}
 \eea
As a result, the gap equation for $m^*_\phi$ becomes
 \begin{multline}
 m_\phi^{*2}=\mu^2_R-{g_4^R\over 4\pi}\bigg(\mu_0+{1\over 2}\sqrt{m_\phi^{*2}-\mu_0^2}\,\arccos
 (\mu_0/ m_\phi^*)\bigg)
 \\
 +{g_6\over 16\pi^2}\bigg(\mu_0+{1\over 2}\sqrt{m_\phi^{*2}-\mu_0^2}\,\arccos
 (\mu_0/ m_\phi^*)\bigg)^2~.
 \label{scalargap2}
 \end{multline}

Solutions of this gap equation describe the stationary points of the effective potential.
In the following we analyze these solutions
and demonstrate that the quenched model exhibits peculiar phase structures.

\section{Phase structure of the model}
To analyze the admissible phases of the model let us derive the
effective potential of the theory at $t\rightarrow\infty$. From \reef{scalaction}, we get,
up to $\bar\lambda$-independent constant,
 \be
 V_{eff}(\bar\rho,m_\phi^{*2})={\mu^2\over 2}\bar\rho+{g_4 \over 4}\bar\rho^2+{g_6 \over 6}\bar\rho^3-{m_\phi^{*2}\,\bar\rho\over 2}
 +{1\over 2} \int_0^{m_\phi^{*2}} dm^2 \int^\Lambda {d^2 p \over (2\pi)^2} \, G_\phi(t,t;p;\mu_0,m)~.
 \ee
Varying this effective potential with respect to $\bar\rho$ and $m_\phi^{*2}$
correctly reproduces the saddle point equations \reef{scalargap}.
Note that $\rho$ is not a dynamical field since it enters only algebraically
into the action \reef{scalaction}. Hence we eliminate it from the effective potential
using the second equation \reef{scalargap}. Replacing the couplings by renormalized ones and further
rescaling them by $\mu_0$ yields
 \begin{multline}
 \tilde V_{eff}(m^2)=-{\tilde\mu_R^2\over 8\pi}\bigg(1+{1\over 2}\sqrt{m^2-1}\,\arccos (1/\sqrt{m^2})\bigg)
 \\
 +{\tilde g_4^R \over 4(4\pi)^2}\bigg(1+{1\over 2}\sqrt{m^2-1}\,\arccos (1/ \sqrt{m^2})\bigg)^2
 \\
  -{g_6 \over 6(4\pi)^3}\bigg(1+{1\over 2}\sqrt{m^2-1}\,\arccos (1/\sqrt{m^2})\bigg)^3
  \\
+{(m^2+2)\sqrt{m^2-1}\arccos(1/\sqrt{m^2})+m^2-\log m^2\over 48\pi}~,
 \label{effpot1}
 \end{multline}
where $\tilde V_{eff},\,\tilde\mu_R,\,\tilde g_4^R$ and $m^2$ denote
respectively the rescaled dimensionless effective potential, the
dimensionless renormalized couplings and asymptotic mass.

\begin{figure}[t!]
\begin{center}
\includegraphics[scale=0.7]{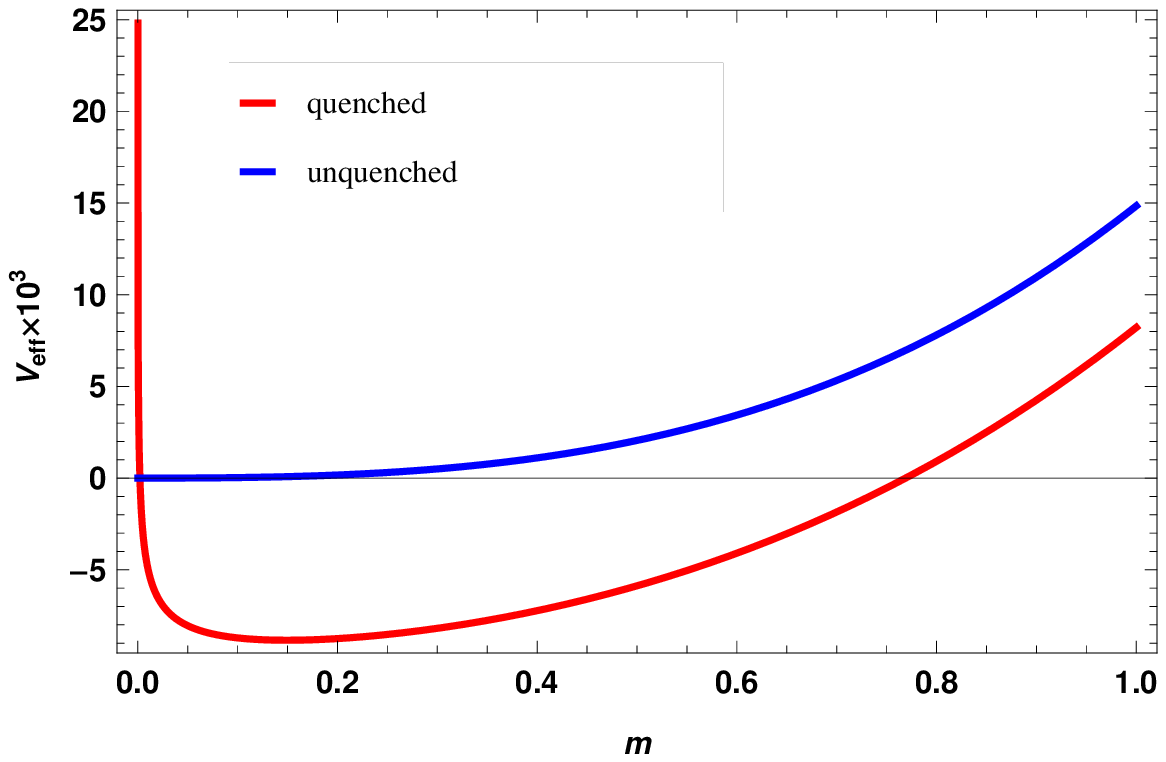}
\caption{Red line (bottom): effective potential (\ref{effpot1}) as a function of $m$ for $g_6=\mu_R=0$ and $\tilde g_4^R=1$.
\newline Blue line (top): effective potential of the $\phi^4$ theory, \ie $g_6=0$, in the absence of quench when $\mu_R=0$ and $\tilde g_4^R=1$.}
\label{fig:phi4}
\end{center}
\end{figure}

Let us briefly discuss the case where both $g_6$ and $\mu_R$ are zero.
This case was considered in \cite{Sotiriadis:2010si}.
The characteristic shape of the effective potential of the quenched theory
in this case is shown in figure \ref{fig:phi4}(red line), and as pointed out in
\cite{Sotiriadis:2010si} a finite mass always emerges at late times
in the presence of interactions.
This should be contrasted with the presence of a global minimum at
$m=0$ in the unquenched theory as shown in figure \ref{fig:phi4}(blue line).
This case already illustrates the main point that we wish to make, namely that the shape of the
effective potential at late times depends on the quench, an event that occurred in the far past.
More interesting and spectacular however is the case when the theory is sitting at the tricritical point,
\ie when $\mu_R=g_4^R=0$.
Expanding effective potential then for large and small values of $m^2$ yields
 \bea
  \tilde V_{eff}(m)&=&{m^3 \over 96}\bigg(1-{g_6\over g_c}  \bigg)+\mathcal{O}(m) \quad\quad \text{if}\quad m>>1~,
  \non
  \tilde V_{eff}(m)&=&-{g_6\over g_c}{\log^3 m\over 12\pi^3} +\mathcal{O}(1)  ~\quad\quad\quad \text{if}\ \quad m<<1~.
 \eea
where $g_c=256$ corresponds to a critical value beyond which the potential is unbounded from below and thus the theory is unstable.

It is remarkable that $g_c$ is larger than the corresponding value in the unquenched case \cite{Bardeen:1983rv}. There, the region of stability is bounded\footnote{The lower bound $g_6\geq0$ is necessary to avoid classical instability. } by $0\leq g_6 \leq (4\pi)^2\equiv g_6^*$.
In particular, previous studies\cite{Townsend:1975kh,Pisarski:1982vz,Bardeen:1983rv} have spelled disaster for the theory in the ultraviolet limit: the $\beta$-function drives the system into a UV fixed point $g_6=192$ which lies beyond
the region of stability.
In contrast, our results indicate that there is a way to circumvent the above conclusion by a quench in the parameters of the system.

\begin{figure}[t]
\begin{center}
\includegraphics[scale=0.5]{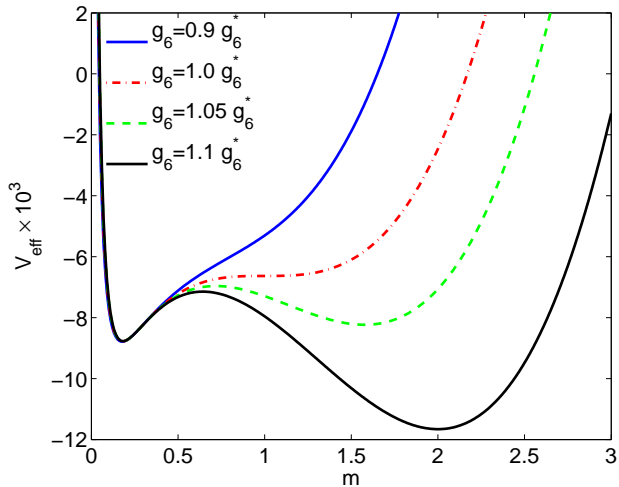}
\caption{Effective potential \reef{effpot1} as a function of $m$ at the tricritical point $\mu_R=g_4^R=0$ in the vicinity of $g_6^* = (4\pi)^2$.}
\label{fig:phases}
\end{center}
\end{figure}

It is instructive to contrast the phase diagrams with those of the unquenched case.
If the coupling constant belongs to the range $0\leq g_6 < (4\pi)^2$ and the system is not quenched, then there is only one admissible conformal phase. Quenching the system in this regime, we explicitly break the conformal invariance by introducing a scale $\mu_0$, and as a result the system resides in the light phase - a unique vacuum state associated  with the minimum of effective potential, see solid blue graph at the top of figure  \ref{fig:phases} which represents a characteristic plot of the effective potential in this case.

On the other hand, if the coupling constant is tuned to the special value $g_6^*$ the unquenched potential becomes flat, and thus a continuum of massive solutions emerges. This continuum is associated with spontaneous breaking of scale invariance and it coexists with the conformal (or massless) phase  which we analyzed before. Quenching the system breaks the scale invariance explicitly and $\mu_0$ singles out a unique vacuum out of the continuum of states. As a result, in the steady state as $t\rightarrow\infty$ we find two admissible vacua shown in figure \ref{fig:phases}.
The heavier phase for this value of the coupling is only meta-stable unless the coupling constant is sufficiently large,
see dashed green and solid black graphs at the bottom of figure \ref{fig:phases}.

Of course, in the unquenched case all phases become unstable for $g_6>g_6^*$ and the system rolls down to infinity in a finite time \cite{Asnin:2009bs}. Here we have showed how the quantum quench may enhance the stability of the system such that for $g_6>g_6^*$ these phases are stable and the escape to infinity is avoided until $g_6$ hits $g_c$.

\section{Concluding remarks.}

To conclude, by studying the late time phase structure of the $\phi^6$ theory
after a quantum quench, we have essentially demonstrated in a specific example
the following: \emph{a dramatic event that
occurred in the far past can have significant effects even in the far
future}.
Not entirely unexpectedly, we find that in the large $N$ limit
the late time physics cannot be described by simple
thermalization with a single effective temperature, which has been noted
in many integrable models. Explicit details are relegated to Appendix \ref{sec:thermal}.
Instead of simply thermalizing,
the quantum quench appears to modify the phase structure, even long after
the system has relaxed and settled into an equilibrium state. From the calculations,
it appears that
this is a generic feature of quantum quenches, and is not specific to the model that
we have studied. In passing, we note also
significant modifications in a supersymmetric
version of this model: contrary to the current model, where new stable
phases are created, we found instability generated by the quench \cite{inprogress}.
This dependence of the phase structure on past events may have
implications in other areas of physics, e.g. in cosmology. It
is therefore
important to determine and perhaps classify, different
time dependent changes in a generic theory that could potentially lead
to drastic modification of late time physics.. In this work we have made extensive use of
techniques developed in \cite{Sotiriadis:2010si}, where it is implicitly assumed that
the system relaxes ultimately to an equilibrium. The authors of \cite{Sotiriadis:2010si} support their claim by explicit
numerical computations and show that their assumption works very well in $\phi^4$ theories in arbitrary dimensions.
In this work we extrapolate their assumption to study the $\phi^6$ theory at the tricritical point.
Further numerical checks of the current model and more
general ones are under way and will appear elsewhere \cite{inprogress}.

\appendix
\section {Absence of thermalization:}
\label{sec:thermal}

A question regarding the late time behavior studied in this work is whether it is effectively thermal. However, despite carrying interaction terms, the model we are studying is in fact integrable in the large $N$ approximation, and as noted in previous works, is not expected to be describable by a single effective temperature
\cite{conjecture}. In the rest of this appendix we
 demonstrate that
the late time physics of our large $N$ vector model is incompatible with simple thermalization leading to a single effective temperature.

If the stationary behavior of the system is thermal, \ie the effective mass $m_\phi^*$ is the thermal mass $m_T$ of the system. One can fix the temperature by matching the gap equation \reef{scalargap} at $t\rightarrow\infty$ with that in the thermal case. As a result, one gets the following relation which determines the inverse temperature $\beta$
\be
\langle\phi^2\rangle_0=\langle\phi^2\rangle_\beta~,
\label{match}
\ee
where subscripts ``$\beta$'' and ``$0$"  indicate that the expectation values are taken in the thermal and $|\Psi_0\rangle$ states respectively.

Having fixed the temperature via \reef{match} (implicitly), we would like to compare
the expectation value of the stress tensor of the thermal state with our quenched state at late times.
 The energy-momentum tensor of the scalar $O(N)$ vector model in Minkowski signature is given by
 \be
 T_{\mu\nu}=\del_\mu\phi\cdot\del_\nu\phi
 -\eta_{\mu\nu}\bigg({1\over 2}\del_\mu\phi\cdot\del^\mu\phi-NV(\phi^2/N)\bigg)
 -{1\over 8}(\del_\mu\del_\nu-\eta_{\mu\nu}\del^2)\phi^2~.
 \ee
where $V(\phi^2/N)$ is the potential\footnote{In our case $V(\phi^2/N)={\mu^2\over 2 N}\phi^2+{g_4\over  4 N^2}(\phi^2)^2+{g_6\over  6 N^3}(\phi^2)^3$.}. The Euclidean form $T_{\mu\nu}^E$ is obtained by flipping the sign of the potential and replacing $\eta_{\mu\nu}$ with $\delta_{\mu\nu}$ in the above expression. In the large $N$ limit the expectation value $\langle V(\phi^2/N)\rangle$ equals $V(\langle\phi^2\rangle/N)$. Hence, provided that \reef{match} holds, we get in the limit $t\rightarrow\infty$
 \be
 \langle T_{00}\rangle_0+\langle T_{00}^E\rangle_T=
 \bigg(\eta_{0\al}\eta_{0\bt}-{1\over 2}\eta_{\al\bt}\bigg)
 \langle\del^\al\phi\cdot\del^\bt\phi\rangle_0
  +\bigg(\delta_{0\al}\delta_{0\bt}-{1\over 2}\delta_{\al\bt}\bigg)\langle\del^\al\phi\cdot\del^\bt \phi\rangle_\beta~.
  \label{enerdif}
 \ee

The left hand side of the above expression should vanish if the system thermalizes since the  zero-zero component of the Euclidean energy-momentum tensor is the minus energy density.
Consider adding $m^{*2}_\phi\phi^2/2$ to the first term on the right hand side. We immediately identify that as $\langle T_{00}\rangle_0$ of a quenched free scalar field whose mass jumps from $\mu_0$ to $m^*_\phi$. On the other hand, subtracting $m^{*2}_\phi\phi^2/2$ from the second term gives the minus thermal energy of the free scalar field with mass $m_\phi^*$. Since the free theory does not thermalize, there is no temperature such that these two terms cancel each other. This proves a mismatch between thermal and quenched physics.

It is also instructive to compare the thermal free energy of the system at the tricritical point with the corresponding effective potential \reef{effpot1}.  In the large $N$ limit the thermal free energy density $f$ can be evaluated in a closed form and is given by
 \begin{multline}
 f/N={g_6\over 6}\({m_T\over 4\pi}+{\log(1-e^{-\beta m_T})\over 2\pi\beta}\)^3-{m_T^3\over 24\pi}-{m_T^2 \log(1-e^{-\beta m_T})\over 4\pi\beta}
 \\
 +\int_0^{m_T^2} { \log(1-e^{-\beta m_T})\over 4\pi\beta}dm_T^2~.
 \end{multline} 

\begin{figure}[t]
\begin{center}
\includegraphics[scale=0.8]{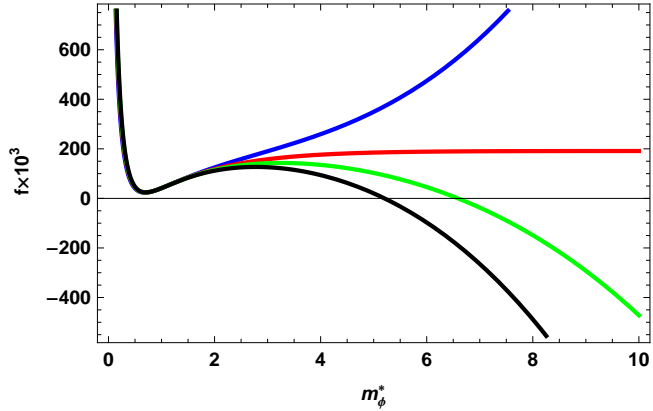}
\caption{The plot of the free energy density $f$ for various values of the coupling
constant $g_6$ as a function of effective mass $m_\phi^*$. $N=1$, the inverse
temperature is set to $\beta=1$ and $g_6=0.9 g_6^*, ~g_6^*, ~1.05 g_6^*,~
1.1 g_6^*$ from top to bottom graph respectively.}
\label{fig:free}
\end{center}
\end{figure}

The typical plots are shown in figure \ref{fig:free}. While for $g_6<g_6^*$ the free energy exhibits certain similarity with the corresponding effective potential, they become manifestly different for $g_6>g_6^*$ and hence thermalization is not expected.

\acknowledgments

We would like to thank J.~Cardy, R.~Leigh, S.~Sachdev, Y.~Shang and especially A.~Buchel, R.\,C.~Myers and S.~Sotiriadis for useful
conversations and correspondence. Research at Perimeter Institute is
supported by the Government of Canada through Industry Canada and by
the Province of Ontario through the Ministry of Research \&
Innovation.  E.S. is partially supported by a National CITA Fellowship.

\end{document}